\documentclass[sigconf,screen,nonacm]{acmart}

\AtBeginDocument{%
  \providecommand\BibTeX{{%
    \normalfont B\kern-0.5em{\scshape i\kern-0.25em b}\kern-0.8em\TeX}}}

\setcopyright{none}

\usepackage{graphicx, color, xcolor, colortbl}
\usepackage{enumitem}

\author{Maryam Molamohammadi}
\authornote{Equal contributions.}
\email{maryam.molamohammadi@mila.quebec}
\affiliation{%
    \institution{Mila - Québec AI Institute}
  \city{Montréal}
  \state{Québec}
  \country{Canada}
}

\author{Afaf Taïk}
\authornotemark[1]
\email{afaf.taik@mila.quebec}
\affiliation{%
    \institution{Mila - Québec AI Institute}
    \institution{Université de Montréal}
    \city{Montréal}
    \state{Québec}
    \country{Canada}
}

\author{Nicolas Le Roux}
\email{nicolas@le-roux.name}
\affiliation{%
    \institution{Microsoft Research}
    \institution{Mila - Québec AI Institute}
    \city{Montréal}
    \state{Québec}
    \country{Canada}
}

\author{Golnoosh Farnadi}
\email{farnadig@mila.quebec}
\affiliation{%
    \institution{Mila - Québec AI Institute}
    \institution{McGill University}
    \institution{Université de Montréal}
    \city{Montréal}
    \state{Québec}
    \country{Canada}
}

\begin{document}

\title{Unraveling the Interconnected Axes of Heterogeneity in Machine Learning for Democratic and Inclusive Advancements}



\begin{abstract}

The growing utilization of machine learning (ML) in decision-making processes raises questions about its benefits to society. In this study, we identify and analyze three axes of heterogeneity that significantly influence the trajectory of ML products. These axes are i) values, culture and regulations, ii) data composition, and iii) resource and infrastructure capacity. We demonstrate how these axes are interdependent and mutually influence one another, emphasizing the need to consider and address them jointly. Unfortunately, the current research landscape falls short in this regard, often failing to adopt a holistic approach. We examine the prevalent practices and methodologies that skew these axes in favor of a selected few, resulting in power concentration, homogenized control, and increased dependency. We discuss how this fragmented study of the three axes poses a significant challenge, leading to an impractical solution space that lacks reflection of real-world scenarios. Addressing these issues is crucial to ensure a more comprehensive understanding of the interconnected nature of society and to foster the democratic and inclusive development of ML systems that are more aligned with real-world complexities and its diverse requirements.

\end{abstract}


\newcommand{\tochange}{\color{green}}
\newcommand{\todo}{\color{red}}
\newcommand{\pointer}{\color{pink}}
\newcommand{\intro}{\color{brown}}
\newcommand{\outline}{\color{brown}}
\newcommand{\magenta}[1]{\textcolor{magenta}{#1}}
\newcommand{\golnoosh}[1]{\magenta{\textsc{Golnoosh:} #1}}
\newcommand{\nicolas}[1]{\magenta{\textsc{Nicolas:} #1}}
\newcommand{\blue}[1]{\textcolor{blue}{#1}}
\newcommand{\violet}[1]{\textcolor{violet}{#1}}
\newcommand{\afaf}[1]{\blue{\textsc{Afaf:} #1}}
\newcommand{\maryam}[1]{\violet{\textsc{Maryam:} #1}}

\maketitle

\section{Introduction}
Machine learning~(ML) products have become pervasive in decision-making processes across various socially significant domains including welfare eligibility ~\cite{fuster2022predictably, liu2018delayed}, public health ~\cite{potash2015predictive}, maltreatment screening ~\cite{vaithianathan2013children}, law enforcement and judicial system ~\cite{koulish2017immigration, angwin2016machine, buolamwini2018gender,o2016weapons}, and labor market and hiring ~\cite{hu2018short,mitchell2018prediction}. However, the current approach to developing ML products for such high-stakes decision-making systems often neglects the diverse needs of stakeholders, specifically end-users. Recent studies have highlighted the potential harms caused by irresponsible design and maintenance of ML products, posing risks to societal progress and the collective advancements made by humanity over generations~\cite{shelby2022sociotechnical}.

This paper emphasizes the importance of designing an ML system that represents both the objective, but also the context in which that system will operate. Driving from the literature and investigating missing connections, we distinguish three main aspects of the design that affect an ML system: i) \textbf{data composition}, ii) \textbf{resource capacity}, and iii) \textbf{values, culture, and regulation}.

\emph{Data composition}  is the most obvious of the three and there is a rich literature on the influence of the training data, its origin and composition, on the output of an ML system. How to generalize to out-of-distribution data and how to be robust to a change in the class mix is an active area of research. \emph{Resource capacity} captures both the amount of compute available to different populations, whether to train these models or to perform inference, and the expertise required to use them. Approaches around distillation, model compression or architecture search all aim to design cheaper and more efficient models. Parameter-free optimization, open-source libraries and online tutorials are different ways to address the expertise gap. Succinctly, how to address the resource gap is another well-studied and understood challenge. The latter of the three is perhaps the least understood. We argue that different \emph{values and culture} influence the design of ML systems in less obvious ways than data and resources do. For instance, what is a desirable objective is a societal decision. One could imagine that a certain context might value accuracy over privacy, but such a choice is not universal.

Not only should all these three axes be considered, they impact each other and need to be considered jointly, as we describe through a few examples. We argue the fragmented discourse, solution space and research areas about each of these three axes, which we call \emph{axes of heterogeneity}, is a problem since it is not reflective of their true nature. The holistic view will equip us to fully understand and plan for harm mitigation interventions, and enable us to answer questions like: How did the common practices and methodologies in the field form by favoring certain ideologies? How cross-context reusability of our models would translate across different value systems, resource capacities, and data heterogeneity? How we can empower end-users and locals to be in charge of design choices and decision making processes?

The concerns arising from fragmented views of these three axes in research are exacerbated when there is an extensive gap, in terms of data availability and resources, between enterprises that are typically developing the systems and a wide range of communities that are impacted by them~\cite{thomas2020problem}. Indeed, despite the intentions of the designers of these systems, they continue embedding normative values of those in power. This design choices and value homogenization is often not well documented, disclosed, or debated in the public domain or through representatives (in the regulatory system)~\cite{kroll2018fallacy,rahwan2018society}. There is enforcement of not-so-invisible control over a variety of ``users'' without fully respecting their rights. This leads to the emergence of the mindset of ``The west and the rest''~\cite{hall1992west} and, especially with the growing excitement and rush toward scaling and designing large models in the ML community, this trend puts tools in the hands of few. This reinforces the old systems of power even further without offering intervention points for the rest of society which results in even more access inequality and higher degrees of colonial AI state~\cite{mohamed2020decolonial,bender2021dangers,crawford2021atlas}. 

Indeed, the same narrative of power, ability to control, and, hegemony would extend beyond data sovereignty into model sovereignty, and reinforce wielded power and dependency with a new facade. Furthermore, since the fragmented discourse about these three axes will include technical methods as well, we describe technical disconnects that are in the current landscape of ML developments. Despite the numerous solutions proposed to address the heterogeneity in terms of data, resources, and values individually, many of them have some dead angles and are not effective in conditions where we experience multiple heterogeneities, which is often the case. Through these discussions, we describe the challenges of democratizing algorithmic design and discuss the limitations of existing technical solutions. By considering the interplay between data composition, resource capacity, and values, culture, and regulation, ML system designers can create context-aware solutions that better address real-world challenges. This paper contributes to the understanding of these aspects and advocates for their comprehensive consideration in ML system design, and calls for new socio-technical frameworks enabling a truly inclusive and sustainable ML development landscape.

\section{Axes of Heterogeneity in Machine Learning}
Building reliable and relevant ML systems is a desirable yet arduous task. Aligned with numerous definitions of an ML model \cite{mitchell2007machine}, we define a \emph{model} as a mathematical representation of a system that can learn from experience and make predictions or decisions based on input data. In this section, we distinguish three axes of heterogeneity that influence the path of every ML model and can cause deviation among seemingly similar ones: \textbf{data heterogeneity}, \textbf{resource heterogeneity} and \textbf{values, culture and regulation heterogeneity}. These three axes of heterogeneity are not independent, and their multifaceted interactions add levels of complexity that reflect real-world scenarios. In this section, we first elaborate on each axes and then discuss their intersections.

\subsection{Data Heterogeneity}
\label{sec:data_het}

Data plays a significant role in the ML pipeline, exerting influence across various stages including data preprocessing, feature engineering, model selection, training, and evaluation. In this context, the first axis of heterogeneity we explore is data heterogeneity.

We can distinguish several underlying causes that will lead to data heterogeneity. It includes data representation, distribution and quality which are heavily influenced by the geo-cultural and institutional context in which it was collected and 
processed. Starting with data representation, data heterogeneity can lead to representation bias, where certain communities or demographics are underrepresented or misrepresented in the data. This can result in biased ML models that fail to capture the nuances and characteristics of these groups, leading to inaccurate predictions and unfair outcomes. Addressing representation bias requires efforts to collect diverse and representative data that includes sufficient samples from all relevant communities and demographics. Furthermore, even the selection of features, attributes or unobserved features that are approximated through observable proxy attributes are heavily subjective processes that include making assumptions \cite{jacobs2021measurement}. These agreements are usually selected by the conductor of the research or common practices in the field that define how to approach the problem at hand given the constraints. Whereby, some decisions are not universal and would affect the nature of data.

Various types of distribution shifts across domains are preventing efficient and reliable adoption of models in different contexts~\cite{hooker2020characterising}. Different types of distribution shifts can include the following descriptions. \emph{Demographic shift} describes a shift in the distribution of the demographics of the protected attributes of the population in the source and target datasets. An illustration can be the presence of more females in the target distribution compared to the source distribution. \emph{Covariate shift} defines a shift in the distribution of the input features among source and target datasets, e.g., a shift in the incomes of loan applicants across geographical regions which impacts the chances of receiving positive outcomes and exacerbates the unfairness of models \cite{ding2021retiring}. \emph{Label and concept shift} identifies a shift in the distribution of the outcomes. \emph{Acquisition/Measurement shift} describes when the relationship between the input variables and proxies differs among the source and target. This can happen when there is different measurement instruments or metric for obtaining or processing input data. For instance, the use of different scanner protocols in medical imaging is a well-studied category of shifts \cite{castro2020causality}. \emph{Manifestation shift} refers to when the causal relationship between an observable attribute, whether reflected in the data or not, and target changes. For example, disease status physically manifests in the anatomy changes between domains \cite{castro2020causality}. This can be a likely scenario in a social environment as individuals have different scales of social settings among different geolocations. \emph{Annotation shift} introduces a situation when annotators' lived experience and relevancy to the task differs, alongside annotation guidelines, which results in similar features being labeled differently\cite{moreno2012unifying, castro2020causality}. \emph{Spurious correlation shift} refers to when the casual relationship among attributes differs from source to target domains\cite{wiles2021fine, beery2018recognition}. For instance, zip code sometimes is used as a proxy for race, whereas this may not be static through time nor is it an adequate proxy in different locations. \emph{Emerged label shift} describes when there are unseen labels in the target data. \emph{Relational identity shift} presents situations where a certain identity group is considered a protected attribute in the source environment but then inherits a different nature in the target. The literature mostly assumes that identity is discrete or predefined, using binary gender or mutually exclusive racial categories \cite{lu2022subverting}. Whereas, identity itself, the degree and quality of inter-group association are dynamic and dependent on the environment, historical and political institutional basis in which people are shaping~\cite{shapiro2010relational, lu2022subverting}. 

In the field of domain adaptation research~\cite{domainadaptation}, there is a significant focus on studying different types of data shifts. However, a common limitation is that these shifts are often considered in isolation, leading to unrealistic assumptions in the development of solutions. In real-world applications, it is more common to encounter multiple types of data shifts occurring concurrently, adding to the complexity of the problem. One particular aspect of data heterogeneity that adds to this complexity is the presence of \emph{temporal shifts}. Temporal shifts refer to dynamic changes that occur in the environment over time, leading to variations in the data. For instance, in the healthcare domain, temporal shifts can arise from changes at the patient level, shifts in medical practices, or modifications in administrative policies and procedures~\cite{otto2021origins}. Similarly, in the context of loan applications, macroeconomic trends and evolving preferences can contribute to variations in loan applicants over time~\cite{rezaei2021robust}.

Lastly, data heterogeneity in terms of quality can arise due to multiple reasons. Firstly, data may be collected from different sources with varying levels of accuracy and reliability. For example, data collected from different communities or demographics may exhibit discrepancies in terms of data quality due to differences in data collection methods or biases in data sampling. Secondly, data quality heterogeneity can result from inconsistencies or errors within the data itself. This can occur due to human errors during data entry or data preprocessing, technical issues during data collection, or inherent noise present in real-world data. Thirdly, the quality of the data can be influenced by contextual factors and evolving conditions. For instance, data collected over time may suffer from concept drift, where the underlying relationships between variables change, rendering the existing data less representative or informative for current ML tasks. ML practitioners often examine and preprocess the data to identify and handle outliers, missing values, and other data quality issues while ignoring the contextual aspects that highly influence on the quality of the data at hand.

It is important to acknowledge that these variances in data heterogeneity are sometimes inevitable and should be respected, as they partly reflect the genuine diversity arising from geo-cultural and institutional differences. As ML aims for generalizability, it becomes imperative to consider these diverse aspects of data heterogeneity and assess the feasibility of achieving such generalization. Moreover, emphasizing the significance of data documentation further strengthens the understanding and transparency of the data used in ML, as highlighted in the concept of data sheets \cite{gebru2021datasheets}.
\subsection{Resource Heterogeneity}\label{resources}
The second axis of heterogeneity is resource heterogeneity. The scarcity of resources can create disparities and inequalities, as only those with sufficient financial or technical means can fully engage in ML development. This exclusion limits the diversity of perspectives and contributions, resulting in potential biases and a lack of representation in the models and systems being developed. Furthermore, the absence of community involvement may lead to ML solutions that do not adequately address the unique needs and challenges faced by marginalized or underrepresented communities.

Throughout the stages of an ML pipeline, various resources are required, including ML and data expertise, domain experts, data curation and processing capabilities, as well as computational and storage resources. Unfortunately, the high costs associated with these resources often make them inaccessible for many communities, hindering their ability to actively participate in the design and implementation of ML pipelines.

One crucial resource is the involvement of multi-disciplinary domain experts who can provide valuable insights and ensure that the problem being addressed is relevant and meaningful. Their expertise is vital in designing tailored research or product methodologies and offering consultations throughout the project.

Data collection is another resource-intensive aspect. It may involve acquiring measurement tools, designing and conducting surveys, or even purchasing data from external sources. In order to capture a comprehensive understanding of the problem space, it is important to include the lived experiences of users, particularly those from under-represented groups. This requires additional domain knowledge and expertise, often obtained through collaborations with local communities or experts who possess a deep understanding of the specific context.

The high cost of ML practitioners or the scarcity of available talent poses an additional challenge for smaller communities seeking to adopt inclusive ML practices. The expertise and skills of ML practitioners are essential for effectively developing and implementing ML models. However, limited availability or high costs can act as barriers, preventing smaller communities from fully embracing inclusive ML adoption.

In addition to the previously mentioned resources, the expertise of social workers plays a crucial role in establishing meaningful connections with communities, formulating surveys in an appropriate manner, and ensuring the safe and inclusive participation of individuals \cite{muller2022social}. An illustrative example is the \emph{Colors of Covid project}, a grassroots community initiative that aimed to gather data on Covid-19 infections among Black, Indigenous, and people of color (BIPOC) communities. The project also collected statistics on food insecurity, job losses, and the overall mental health status of participants. It is important to recognize that different cultures and communities may have unique perspectives and experiences, which can influence their responses to survey questions. For instance, in the \emph{Colors of Covid project} people responded differently to the question "Have you ever experienced trauma?" and some cultures are more reserved or have different ways of expressing such experiences \cite{garcia2021color}. Social workers, with their cultural competence and sensitivity, can help ensure that survey questions are framed in a way that respects and acknowledges diverse perspectives, allowing for more accurate and meaningful data collection.

In the context of data, small communities or low-resource institutions often face challenges related to data scarcity and limited data digitization capabilities, which can have a detrimental effect on the performance of ML models. Additionally, in supervised tasks, data annotation poses another obstacle that requires the expertise of domain experts to develop annotation guidelines and access to a pool of qualified annotators.

Finally, the availability of infrastructure, including private servers and cloud services for data storage, model training, deployment, and maintenance, presents another challenge for small communities or companies. These infrastructures often come with significant costs, making them financially inaccessible for these entities. Training a reliable ML model alone can incur substantial computation costs, which are completely out of reach for small communities or companies. To put this into perspective, observations from 2018 indicated that the amount of computing power used to train the largest deep learning models had increased by a staggering 300,000 times in just six years \cite{amodei2018ai}. This trend, combined with the current enthusiasm and focus on scaling and designing large models in the ML community, further widens the gap between resourceful and powerful entities and the rest. 

Addressing the resource gap is crucial to foster inclusivity and empowering communities to actively participate in ML development. Efforts should be made to provide accessible resources, such as affordable computing infrastructure, open-source tools and libraries, and educational programs that enhance ML and data literacy. Collaboration and knowledge-sharing initiatives can also facilitate the transfer of expertise and empower communities to participate in building and deploying ML models that align with diverse requirements.

\subsection{Values, culture and regulation Heterogeneity}
\label{values} 

Data heterogeneity has been extensively studied in the literature, focusing on the influence of training data on system outputs. Resource heterogeneity involves available compute resources and expertise, with efforts to develop cheaper and more efficient models. In contrast, the influence of the third axes of heterogeneity, values, culture, and regulation, on ML system design is less apparent compared to data and resource considerations. Hence in this section we discuss the third axes of heterogeneity that we refer to as value Heterogeneity.

The ultimate goal of developing ML tools is to enhance the quality of life for all individuals. However, it is crucial to acknowledge and embrace the diverse and pluralistic nature of our global society, which encompasses a wide range of ideologies and perspectives. Different values, cultures and regulations can mark and set preferences and constraints in different stages of ML research and development. 
The influenced areas include but are not limited to research interests, methodology, problem formulation and abstraction \cite{selbst2019fairness,passi2019problem, jacobs2021measurement}, data collection and treatment \cite{miceli2020between}, and ML model design choices~\cite{hooker2021moving,mitchell2019model}. For instance, the research questions that are chosen and supported, as well as the processes employed in research, can reflect specific values. This can manifest in the preference for large models over low-resource alternatives or the promotion of collaborative paradigms. Moreover, the structures and policies surrounding the deployment and maintenance of ML systems are also influenced by cultural and regulatory factors.

Problem formulation encompasses translation and making goals amenable \cite{passi2019problem}. The process concludes a set of abstractions that are usually decided by the conductor of the research or common practices in the field on how to approach the problem at hand given the constraints. The overarching goal is under influence of context, scale, drive that ranges from social welfare to business drive, values, culture and regulation \cite{mitchell2018prediction}. 

Regarding ML design choices, the primary emphasis in the development of ML lies in the efficacy and generalizability of the algorithmic process itself. Even this focus, is often skewed by the pioneers of the field and their values. The ethos of accumulation and maximizing short-term benefits explain the choice of accuracy as the common performance metric. These choices often reflect individualistic drives or reductive ways of evaluation that impose certain ideals in model development. Whereas, the attitudes towards civic duty and common good differ among people with different cultural backgrounds \cite{waldron2000cultural}.

Algorithmic fairness has emerged as a promising avenue to address bias and discrimination by incorporating different values and perspectives. However, it is important to recognize the limitations of universal fairness definitions. Fairness is a nuanced concept influenced by social context and diverse viewpoints on fundamental values in life. It should not be constrained to a single perception or standard \cite{sandel1998liberalism}. Unfortunately, current approaches to algorithmic fairness often rely on simplified definitions rooted in a eurocentric account of morality, disregarding alternative moral philosophies and the experiences of diverse communities. Fairness should be understood as a subjective perception shaped by past interactions and cultural values that vary across societies \cite{sandel1998liberalism, grgic2020dimensions}. Embracing this perspective can foster a more inclusive and contextually grounded approach to algorithmic fairness.

In addition, fairness in machine learning is a complex and dynamic concept that goes beyond the technical challenges of defining a universally satisfactory notion. It is influenced by various factors, including values, culture, and regulations, which vary across different contexts and evolve over time. Recognizing the dynamic nature of fairness is crucial when designing ML models, as the understanding and interpretation of fairness can change. It requires ongoing evaluation, adaptation, and responsiveness to societal changes and emerging fairness considerations. 

Moreover, addressing different regulations on anti-discrimination laws and data protection laws around the world poses significant challenges in the development of ML pipelines. Various countries have enacted legislation to prevent discriminatory practices and protect individuals' privacy rights. However, these regulations can vary significantly in scope, definitions, and requirements across jurisdictions. Adhering to these laws requires ML practitioners to navigate complex legal frameworks, understand jurisdictional boundaries, and ensure compliance throughout the entire pipeline. Additionally, data processing activities, such as data collection, storage, and sharing, need to align with the principles of data protection, which can include obtaining informed consent, anonymization, and secure data handling. 

It is important to note that while addressing both fairness and privacy is essential in designing a democratic and inclusive ML model, there is often a clear trade-off between privacy and fairness in practice. Differential privacy, the defacto technique for privacy in ML, involves adding noise to the data or the model. This disproportionately affects minority groups in the datasets, as they are either more exposed to privacy threats \cite{chang2021privacy} or too much noise is added to their datapoints which leads to inaccurate or unfair predictions \cite{sanyal2022unfair}. Moreover, privacy in data collection itself is often regarded as a monolithic concept where people do not experience privacy equally and it is different among different identity groups \cite{marwick2018understanding}. This can dictate the kind of data that can be collected, as well as what applications can be adopted and deployed. Furthermore, on top of fairness, values and culture variances affect the expectations of privacy, and the relationship individuals have with privacy can change considerably depending on the situation and social dynamics at play, e.g., in difficult physical health situations and with the risk of losing insurance, people's priority and sense of autonomy differ~\cite{marwick2018understanding}.

Legal and regulatory frameworks in certain jurisdictions have recognized the significance of explainability in automated decision-making systems. The General Data Protection Regulation (GDPR) in the European Union, for example, includes provisions that grant individuals the right to receive an explanation when automated decisions are made about them. These requirements aim to ensure that individuals have the opportunity to comprehend the factors influencing decisions that have an impact on their lives, particularly in sensitive or high-stakes domains. By enforcing explainability, the law aims to foster transparency and prevent the occurrence of unfair or discriminatory practices. However, the challenge lies in determining how the system's output should be explained to end-users, as it can vary depending on the context, and the expectation of the end-users. In healthcare, the explainability of algorithms becomes crucial as it directly impacts patient well-being and safety. Doctors, patients, and regulatory bodies may expect a high degree of transparency in understanding how a diagnosis or treatment recommendation was reached. In this context, detailed explanations that highlight the key factors, medical evidence, and reasoning behind the decisions are crucial for building trust and ensuring accountability. Healthcare professionals, who possess specialized knowledge and expertise, may require detailed explanations that delve into the technical aspects of the decision-making process. They seek insights into the underlying algorithms, medical evidence, and the reasoning behind the recommendation to ensure clinical justifications and evaluate the system's reliability. On the other hand, patients may have different expectations and information needs. They may prioritize receiving explanations in a more understandable and patient-friendly manner, focusing on the potential benefits, risks, and alternatives of the recommended intervention.

Lastly, the trade-off between fairness, privacy, and explainability requirements mandated by law presents a challenge in designing ML models. Techniques like differential privacy, which aim to protect privacy, can impact the ability to provide detailed and comprehensive explanations of the model's outputs~(Saifullah et al., 2022). This creates a delicate balance that communities must navigate in order to address the competing demands and prioritize their objectives. It requires careful consideration and reconciliation of privacy, fairness, and explainability within the specific context of each community, taking into account their unique vision, policies, and regulatory frameworks. 

\subsection{Intersectionality of the Heterogeneity Axes}
\label{discussion}

To establish a realistic and effective design of ML pipelines, it is crucial to consider the simultaneous presence of all the aforementioned heterogeneity axes. The convergence of shifts in data, diversity in resources, and conflicts in values is a common occurrence in real-world scenarios. This necessitates a comprehensive understanding of the ML pipeline at both a system-level and a contextual level. By adopting a holistic yet specialized perspective, we can lay the groundwork for designing ML pipelines that are well-suited to their intended contexts. Below we showcase the intersectionality of the three axes through a few examples.


\subsubsection{\textbf{Data and value heterogeneities}}
\paragraph{Affirmative actions: Not every protected group needs the same remedy!:} 
Beyond the geo-cultural lens, even in a Western account of morality, justice is not always about equality. Justice is relative to social structures. However, the current definitions of algorithmic group fairness such as equal opportunity and demographic parity are based on equalizing model performance metrics (e.g., accuracy, true positive rates) among different demographic groups \cite{zhang2022affirmative}. 
 
Affirmative-action policies vary from region to region and might take different forms, from targeted encouragement to hard quotas. Affirmative actions in the form of quotas in particular are fundamentally incompatible with other definitions in algorithmic fairness \cite{ho2020affirmative}. As an example, affirmative actions and reconciliation strategies towards indigenous communities in North America cannot be directly replicated in other places in the world, not only because of the historical context but also the demographic makeup of the populations \cite{schumann2019transfer}. 
 



\paragraph{High-stake events:} Another compelling scenario that necessitates redefining fairness and inclusivity arises in the context of extreme events such as war and natural disasters. These events bring about significant social transformations, including the influx of refugees into host countries or regions, consequently altering their composition and demographic distribution. In response, host countries often endeavor to develop new policies and programs aimed at promoting the integration of these newcomers into areas such as schools, employment, and housing, as highlighted by Brännström et al. \cite{brannstrom2018lived}.

Given the high-stakes nature of these changes in data distribution and the dynamic concept of fairness, it becomes crucial to develop models that can adapt to such transformative circumstances when incorporating algorithmic decision making. It is imperative to ensure that the decision-making processes remain responsive and sensitive to the evolving landscape and diverse needs emerging from these extreme events. This calls for the construction of models that can flexibly redefine fairness definitions to address the unique challenges presented by these contexts.


\paragraph{Representativeness:} In the pursuit of achieving specific representation in data, pre-processing fairness notions often involve implementing corrective measures. However, it is essential to recognize that the types of interventions and corrective measures employed are strongly influenced by the context and culture in which they are applied. Representativeness, a concept studied across various fields such as statistics, politics, and machine learning, inherently reflects the values embedded within its definitions \cite{chasalow2021representativeness}.

This realization highlights a potential tension between the desire for generalizability and the context-specific nature of representativeness. While the goal is to develop fair and inclusive models that can be applied broadly, it becomes apparent that achieving representativeness requires a deeper understanding of local knowledge and engagement. 

\paragraph{Problem understanding:} In practical ML applications, the choice of metrics and approaches to problem-solving often rely on established practices or metrics defined by the researchers. However, it is important to recognize that some of these metrics may not be relevant or may require additional context to accurately address the problem at hand. In real-world scenarios, local domain expertise and statistical knowledge play a crucial role in formulating problems, determining appropriate data collection methods, and designing ML pipelines. The definition of protected groups, for example, can vary depending on the specific context, and local knowledge is needed to identify the relevant demographic attributes and prioritize fairness evaluations accordingly \cite{moreidentities, schumann2019transfer}.



\subsubsection{\textbf{Data and resource heterogeneities}}

\paragraph{Proxies and Relational Mismatch} In the realm of use-cases, the availability of required attributes cannot always be guaranteed. As a result, it becomes crucial to possess additional knowledge in order to devise appropriate solutions, such as employing suitable proxies. A notable example is the utilization of zip code as a proxy for race, as demonstrated by \cite{noauthor_using_2023}. While these proxies may be adequate in certain locations, their generalizability to other parts of the world becomes questionable. This relational mismatch in the selection of attributes can give rise to problems when causal reasoning about the issue is carried out without local and domain expertise. Moreover, the choice of assigned proxies can overlook the historical context. Research conducted by Obermeyer et al. \cite{obermeyer2019dissecting} revealed that despite the belief that healthcare cost serves as an effective proxy for needed care based on certain measures of predictive accuracy, this assumption leads to significant racial biases. Consequently, the historically unequal access to care among different subgroups within the US healthcare system is disregarded. It is imperative to acknowledge these challenges and limitations in the use of proxies, as they can have far-reaching consequences on the accuracy and fairness of decision-making processes. 


\paragraph{Problem manifestation} Standardized frameworks are not fully compatible with various geo-cultural contexts, as certain phenomena manifest differently in different domains. As an example, in disease diagnosis and medical imaging, a challenging scenario is that under which the way anti-causal prediction targets, e.g. disease status, physically manifest in the anatomy changes between domains. This cannot be corrected without strong parametric domain-specific assumptions on the nature of these differences \cite{castro2020causality}. Attempting to impose a one-size-fits-all approach without accounting for these variations can lead to inaccurate or inadequate results. It becomes imperative to consider domain-specific knowledge and assumptions to effectively address these challenges and enhance the compatibility of standardized frameworks across different contexts.


\subsubsection{\textbf{Data, resource, and value heterogeneities}}

\paragraph{Annotation mismatch:} In the context of supervised learning, different variables with regard to resource capacities and values affect the quality and reliability of datasets' annotation. labeling is an essential yet expensive operation, thus often offloaded to crowdsourcing platforms as a means to reduce its cost. The crowdsourcing platforms offer low compensation and no recognition for the work of annotators. Undermining data labeling work and "invisible workers" contribution \cite{irani2019justice} as opposed to considering it a collaboration also can influence the annotation process and quality of data. Furthermore, a lack of or difficulty to come to a consensus towards standardized annotation agreements, as well as the lack of incentives can translate into neglecting the geo-cultural context, leading to biased and inaccurate labeling. In fact, the work of annotators naturally is informed by their personal values, as well as their cultural and historical backgrounds.

Based on the annotator's knowledge and their lived experience, data would be likely labeled by association or proximity. As an example of annotation mismatch, a study showed that the labels of brides and grooms photos from different countries in the Open Images dataset show Eurocentric bias, as the bride-groom images from Ethiopia and Pakistan are not labeled as consistently as images from the United States and Australia \cite{shankar2017no}. Another example can be found in natural language processing applications, where the heterogeneity of values and cultures encompasses various aspects of communication, such as word choices and connotations, style, intonation, geo-cultural norms, and varying hate speech regulations and their treatment approaches \cite{sap-etal-2019-risk}. In these examples, if there were no constraints on resource capacity, it would have been possible to engage domain experts for annotation tasks or consult with them to establish guidelines for the annotation task. 

\section{Wielded Power Asymmetry, Control and Dependency}

The discussion on democratic advancements in ML is incomplete if it does not consider people, different stakeholders, power dynamics that ML systems are created based on and further reinforcing. The interaction of the three axes of heterogeneity would result in power imbalance among different actors. 
The unequal distribution of resources, data, and influence, among different entities creates a power hierarchy. In this section, we discuss the power concentration in the field and its impact on control, homogenization and trust for less-endowed entities.



\paragraph{\textbf{Wielded power:}} Historically, after humans adapted to the sedentary life, power was created based on accessing knowledge and tools or leveraging various resources simultaneously \cite{diamond2001guns}. ML is no exception;
the problem is that currently, because of the extreme resource imbalance, all three axes weigh in favor of the already advantaged entities that are typically advancing ML systems. More resourceful, hence powerful, entities are well ahead through their access to data, computational resources, funding, and expertise. Indeed, the unequal distribution of resources, data, influence, and their interactions, created power concentration. Large language models, where scaling has been chosen as the design strategy, are a clear example of this power imbalance. Large datasets and models require large compute and storage capacities, which only resourceful entities have. These new trends in research and industry lead to the retention and wielding of power, thus reinforcing dependence relations on certain entities. In fact, if the scaling trend persists, resourceful entities will have the power and decision authority of letting others to the degrees of access or involvement. Therefore, the rest of society is left with little to no control over the decision-making process, interventions, or revising the embedded assumptions and choices. The legitimacy of such power is highly questionable in a democratic society \cite{birhane2022power,maaslegitimacy, thomas2020problem}.



\paragraph{\textbf{Homogenization and control:}}

Throughout history, philosophers have tried to detach distributive justice from moral desert and values, with some considering this as a major divide between modern and ancient thoughts~\cite{sandel1998liberalism}. Morality and values are mainly related to the collective lived experiences of communities, and vary depending on the context. Therefore, ML practitioners should not have the autonomy of imposing a particular definition by hard-coding it into ML algorithms, as such practice does not leave much room for freedom of choice and does flatten and eradicate the ontological and epistemological differences. 

Having the power enables hardcoding certain normative values in ML systems, the ethos of generalizability and prescribed data processing that can lead to the inadvertent or deliberate homogenization of society based on specific preferences. This homogenization is often not disclosed or debated. This control suppresses the unique qualities and perspectives of different communities. In this climate, few have the ability to control the narrative, the data, and the model's design. Given the limited public literacy on ML systems, individuals may remain unaware of the level of control and homogenization in their daily life. The legitimacy of such power is highly questionable in a democratic society \cite{birhane2022power,maaslegitimacy} especially in light of uncertain accountability.

\paragraph{\textit{\textbf{Power dynamics and trust:}}}
Powerful entities remain in control as long as they are trusted, however they might abuse their power and manipulate this trust. Under these assumptions, accountability and auditing are the resistance of the less powerful. In fact, trust is a way of delegating power to certain entities \cite{participatory_urban}. The entities could be a handful of experts or community representatives, and subsequently, an ML model. Thus, trust is a critical factor in deciding whether or not to deploy a model, and if deployed, how to rely on its outputs for critical applications, especially in terms of locally defined values such as fairness and privacy. As a result, resourceful entities usually aim to present their models as trustworthy. 

Trust should be earned and maintained, particularly among unequal entities. 
In fact, the powerful remain in power because they gained trust, and without regulation that trust is misused. Indeed, resourceful entities will take advantage of the absence of regulations in certain countries, and loopholes in policies in others, to build tools without any considerations for privacy, fairness and transparency. 
For instance, to make their ML products more appealing in the market, some companies might engage in what is known as 'fairness washing' \cite{fair_wash}. This involves conducting fairness analyses on metrics of their own choosing, which enables them to claim a level of fairness as they continue to use incomplete and ineffective automated tools, while also avoiding accusations of unethical behavior. Another similar example is 'privacy washing' \cite{domingo2021limits}, where companies claim to use differential privacy, while only optimizing for utility. In fact, using differential privacy requires adding noise to data points to protect privacy, which comes at the price of a loss of utility. By adding very little noise to data, they can still benefit from the claim that they use privacy to gain end-users' trust. 
Consequently, giving an entity the power to deploy a model should not also mean giving up the power to regulate and audit its usage.

\section{Technical Disconnect}

In this section, we portray the current landscape of ML techniques used to treat the three axes of heterogeneity. The discussion also includes the level of control that each technique offers to different stakeholders, and how trust can be established among the stakeholders under varying power dynamics.

It is worth noting that, one of the major assumptions that is usually the case with regard to the techniques is that values are considered direct and tangible means for ML pipeline design. For instance, algorithmic fairness notions are usually used during training as a constraint on an optimization problem, as the regularizer in the loss function, or as an objective itself in a multi-objective setting \cite{zafar2017fairness, choi2020learning, martinez2020minimax, agarwal2018reductions}. Whereas, as mentioned in the previous sections, it does not capture the full scope of what we mean by values. Even if we accept this assumption, our analysis shows different stakeholders especially end-users, communities, and small entities have little to no control over how to design, stay included, intervene, or effectively adapt the models to their needs, environment, and resources. 
\subsection{Data Sharing as Means for Inclusion}

Representation in the training data is among the widely discussed issues of ML models \cite{shankar2017classification}. A possible solution to being included and represented in the training dataset is data-sharing, either by making open datasets \cite{kostkova2016owns} or sharing data in a privacy-preserving manner \cite{jin2019review} with more resourceful entities, using techniques such as encryption \cite{patranabis2016provably}, sharing aggregated data \cite{XIE2011576}, synthetic data \cite{synthetic} or purely sharing data in crisis \cite{peiffer2020machine}.
This means that different communities have to give up their data autonomy and agency in order to be included in the development process of large ML models.

Firstly, if the data is shared with resourceful entities, this hidden contract on the surface may appear volunteer-based and reciprocal to some extent, whereas it has many similarities with colonial-era agreements where the benefits were heavily skewed in favor of the powerful and is a mean of exploitation and a modern extension of colonialism \cite{couldry2019data}. 


Additionally, this scenario also assumes that the powerful entities are trustworthy. This means that they will respect the privacy of the data and use it with the benefit of such communities in mind. For instance, the Maori community in New Zealand collected data and established data
sovereignty protocols \cite{rainie2019indigenous} to ensure that technologies using such data benefit the Maori people first and foremost. Sharing such data with commercial and resourceful entities would mean that such entities will develop models with the benefit of the people in mind. Such utopian assumptions towards near-monopolies entities, would also require that powerful companies would include all the affected stakeholders in the scoping, design and implementation of such solutions. Unfortunately, this is not the case. An example is datasets collected from African countries with the aim to develop models to help fight poverty and other local challenges, conversations around said challenges were often conducted by non-African stakeholders \cite{abebe2021narratives}. Consequently, the affected communities have little to no sovereignty over their data and no control over the models developed using them.

Yet, we can think of alternative scenarios of data-sharing where power dynamics are different. For instance, on more even grounds, several partners with similar priorities and objectives might pool their data together and achieve agreements towards collective ownership and privacy of their data, as well as the intended use and the training of an ML model that benefits all of them. This collaboration will allow them to benefit from the advances in ML and can narrow the digital divide that the need for resources created. As an example, a group of local farmers can come consensus on what they need for optimizing the prosperity of their fields. The challenge is that governing such a system is not always straightforward.

\textbf{\textit{Discussion:}} On top of what we mentioned about giving away data autonomy and agency, data sharing does not necessarily guarantee the desired benefits, particularly in the case of large-scale models. Notably, recent examples, such as the GPT models, have demonstrated performance deficiencies on benchmark datasets, despite the inclusion of these datasets in the training process. This observation raises an essential point about the limitations of relying solely on data inclusion as a solution to ensure fairness and improved performance~\cite{carlini2022quantifying}. It highlights the need for a more comprehensive approach that goes beyond the mere presence of relevant data. Factors such as model architecture, algorithmic design, and the nature of the data itself play crucial roles in determining the model’s overall performance, including its fairness characteristics.

In addition, it is important to consider the impact of data size and distribution on model behavior. Particularly when dealing with small datasets, there is a risk of encountering a phenomenon known as memorization. Memorization occurs when a model simply memorizes the training data instead of learning meaningful patterns and relationships. This can become a significant issue if the distribution of the small dataset differs significantly from the distribution of the majority of data or if it is not representative of the diverse population the model will encounter during deployment. When a model memorizes the training data, it may exhibit high accuracy on the training set itself. However, its generalization and performance on unseen data, particularly from underrepresented groups or diverse contexts, can be severely compromised. This can result in biased or unfair outcomes, as the model fails to capture the complexity and diversity present in the real-world data. In fact, as models continue to scale in size and complexity, it is anticipated that the problem of memorization will become even more pronounced unless deliberate measures are taken to address it.


\subsection{Large Pretrained Models with Postprocessing and Adaptation}

Large pretrained models have demonstrated multi-task \cite{radford2019language} and few-shot \cite{brown2020language} learning abilities, however their training and development require a high amount of compute and storage capacity \cite{bommasani2021opportunities}.
One possibility would be a scenario where resourceful entities are willing to share their trained models and their documentation with smaller communities, which would then be used locally with some adjustments to the fairness definition (e.g., pre-processing and post-processing techniques), data distribution (e.g., transfer learning, local fine-tuning), or resources (e.g., compression, knowledge distillation). 
However, considering the fast-paced and competitive culture of ML research, uncertain accountability, and overall lack of incentives currently these models are close-sourcing. Thus, the narrative of developing large pre-trained models is already distorted. Furthermore, retraining the entire models locally by less-endowed communities might not be feasible. In this scenario, at best, we can hope that such powerful entities would include different values and data compositions during the training phase, through techniques such as multi-task learning.



\textit{Multi-task learning} is a learning paradigm in which multiple tasks are jointly learned by a shared model \cite{crawshaw2020multi}. While the task definition is dependent on the context, in the standard multi-task learning setup tasks are assumed to be defined on the same data distribution, but with different objectives \cite{caruana1997multitask}. We can consider adding different values (e.g., fairness notions or privacy expectations) as different tasks in a multi-task learning setup. However, this approach is not readily applicable to problems with conflicting tasks such as fairness notions. The impossibility theorem in fairness \cite{saravanakumar2020impossibility} states that the common statistical group fairness notions cannot be satisfied simultaneously. 

The simplest option in this case is to use \textit{post-processing} techniques, which involve adjusting the outputs of a model in order to mitigate any biases or disparities that may be present in the predictions. Examples include thresholding \cite{iosifidis2019fae} and calibration \cite{hebert2018multicalibration}. These adjustments allow to regain some level of control over the outcomes with low computational cost, however they are designed for applications that supposed a human-in-the-loop or some domain expertise and might not be always applicable \cite{caton2020fairness}. Additionally, modifying the model's outputs might not always guarantee the fairness of the process or might clash with another requirement of trustworthy ML, which is explainability \cite{lepri2018fair}. A similar concern about model interpretability can be raised when retraining the model and applying \textit{pre-processing} techniques to the data, especially approaches based on applying perturbations and transformations. \textit{In-processing approaches} on the other hand can be used for fine-tuning the model without affecting the interpretability of the model, but it requires that the optimization function is easily accessible and modifiable. Communities can also potentially fine-tune large pretrained models to better align them with their unique data. This can be achieved using domain adaptation techniques such as transfer learning \cite{zhuang2020comprehensive, pan2010survey}.  

\textit{Transfer learning} aims to leverage knowledge from a source domain to improve and accelerate learning on a target domain. The goal is to obtain transferrable features that can be used for various downstream tasks, thus it can be particularly useful for small communities with limited data and computation. However, the transferred knowledge does not always bring a positive impact on new tasks. If there is little in common or a goal mismatch between the domains, knowledge transfer could be unsuccessful or even hinder learning on the target task, that is the performance and efficiency are actually worse than training from scratch on the target domain \cite{wang2019characterizing}.


Adapting to the local environment can also mean adapting to the local resources. In addition to compression techniques (e.g., pruning, quantization), \textit{knowledge distillation} can be used to adapt to the local resources while leveraging the knowledge of the larger model. Knowledge distillation aims to effectively learn a small student model from a large teacher model \cite{gou2021knowledge}, thus allowing both the transfer of knowledge and adapting to limited resources. This is achieved by extracting potent knowledge from the teacher and using it to guide the training of the student. Facing data and embedded value heterogeneity, one can consider small or new communities' tasks as student tasks in knowledge distillation. It has been recently shown that using knowledge distillation through the usage of soft-labels can help achieve a better fairness-accuracy trade-off for several tasks and fairness notions \cite{chai2022fairness}. Nonetheless, a deep understanding of generalization and capacity of knowledge distillation, especially how to measure the quality of knowledge or the quality of the teacher–student architecture, particularly on fairness notions, is still limited \cite{gou2021knowledge}.

\textbf{\textit{Discussion:}} 
Accessibility and reusability are widely advocated principles towards democratizing ML. However, from the feasibility perspective even if the pretrained model is elaborate, there are limits to what finetuning can do. Importantly, the effect of finetuning in the specific case of large pretrained models is not yet fully studied and it is unclear to what extent a large pretrained model can be adapted to novel downstream tasks \cite{tran2022plex}. We can define finetuning capacity as the ability of a pretrained model to efficiently adapt to new tasks; the question then becomes how to increase the finetuning capacity and how to determine whether finetuning of a large pretrained model is cost-efficient. Additionally, how defining and estimating the cost of adaptation for a new domain is not a straightforward task. While recent works \cite{ding2023parameter, tran2022plex} have tackled this important adaptation problem, the aforementioned questions call for further research to develop better context-dependent evaluation metrics that can clear the ambiguity of the adaptation criteria, ease of computation, and efficiency of the adaptation process. 

Moreover, even regardless of the feasibility and evaluation question, the implementation of these techniques reinforces the need for a resourceful entity to build the initiator model in the first place. It feeds existing power structures, widening the gap between those who have access to these tools and those who do not, and perpetuating power imbalances. To address these issues, it is crucial to prioritize the development of more inclusive and participatory approaches to ML design that account for the diverse perspectives and interests of all stakeholders.


\subsection{Collaborative Training}
One of the approaches towards inclusion is to opt for collaborative training, in which different entities (e.g., institutions, communities, small enterprises) can collaborate with each other through techniques such as federated learning and split learning.  


\emph{Federated Learning} (FL) consists of multiple distributed collaborators that try to solve a common problem under the coordination of a central aggregator while keeping data local \cite{mcmahan2017communication, zhang2021survey, kairouz2021advances}. This gives collaborators more control over their data, and allows very low-resourced collaborators to participate. The main assumption is that collaborators agree on model architecture, and then the central model is built on the simple aggregation of collaborators learning parameters. Therefore, the final model does not necessarily guarantee the best performance for all collaborators especially in light of their data and embedded value heterogeneity.  

FL can allow entities with similar expectations and objectives to train a shared model \cite{chu2021fedfair}. Additionally, it can be used in concurrence with other techniques, such as multi-task learning \cite{smith2017federated,marfoq2021federated}, meta-learning \cite{fallah2020personalized}, and rank learning \cite{mozaffari2021frl}, to enable collaborative training on different tasks. Moreover, while the premise of FL is to avoid sharing the local data, communicating the model updates throughout the training process with the aggregator can reveal sensitive information, but these privacy-preserving aspects of FL can be further enhanced using other techniques such as differential privacy \cite{wei2020federated} and secure multi-party computation for aggregation \cite{bonawitz2016practical}. 

Understanding and balancing the aforementioned concepts and their respective trade-offs, both theoretically and empirically, is challenging. Using FL, one has control over its data and model parameters sharing (collaborators can choose if they like to partially participate through sketched or structured updates \cite{mcmahan2017communication}), but it is not yet guaranteed that FL frameworks can accommodate different data distributions and orthogonal objectives (e.g., different fairness notions) while maintaining efficient communication.




\emph{Split learning} \cite{vepakomma2018split} is another collaborative paradigm in which a deep neural network is split into multiple sections, each of which is trained by a different collaborator. The network is trained by transferring the weights of the last layer, namely the cut layer, of each section to the next section while the data remains private and local. The gradients are backpropagated at the server from its last layer until the cut layer. Only the gradients at the cut layer are sent back to each collaborator. The rest of the backpropagation will be completed locally. There are a few variations of split learning which allow keeping the labels of different collaborators private as well. Generally, split learning enables collaboration under resource heterogeneity since it includes a facilitator server. On the other hand, from an ownership perspective, the server provider ultimately will keep part of the network parameters required for inference which could be a concerning factor.




\textbf{\textit{Discussion:}}
Collaborative training approaches for ML models have the potential to revolutionize the way models are trained and deployed. Collaborative paradigms can provide a better fit for the socio-technical account of ML systems. By control and privacy that they offer over local data, these approaches can ensure that sensitive information remains confidential while still allowing for effective training. Additionally, collaborative training can enable resource-limited collaborators to take part, providing a more inclusive approach to ML. Furthermore, collaboration could balance the power concentration that currently exists by narrowing the gap and empowering locals, communities and small entities to take the authority over their models and meaningfully participate in the design process. 

However, the status quo of ML is that it is a highly specialized field, where only a select few practitioners have the expertise and resources (compute and data). This means that participating in collaborative training would still require certain access to resources. Furthermore, there is still much work to be done in order to fully understand and operationalize the capabilities of collaborative training techniques. One key area that requires attention is the analysis of values heterogeneity, especially when different entities have different regulations, which can highly impact the feasibility and the performance of collaborative training approaches. Additionally, the evaluation of whether and to what extent the collaboration is cost-efficient and effective for collaborators is not straightforward.

To ensure effective and sustainable collaborative training, it is crucial to establish clear guidelines or contracts that define the terms. This will help collaborators make informed decisions about whether or not to participate in a collaborative training project, and foster a culture of trust and transparency among them. Trust building in the context of collaborative paradigms is subject to a different kind of power and control dynamic. In this case, collaborators are on a more even ground. In FL for instance, peers would need to trust others' models being built on good quality data, and that the models are adequately trained \cite{mothukuri2021survey}. Mechanisms for accountability to ensure the quality of the shared models have been proposed such as using blockchain and oracles \cite{bao2019flchain}, alongside peer-to-peer evaluation of the model updates to ensure model convergence \cite{roy2019braintorrent}. Such approaches still need to be extended to other aspects of trustworthy ML, such as fairness and privacy, to ensure an alignment with the different values and regulations.



\section{Conclusion}
Developing democratic and inclusive ML solutions require a simultaneous consideration of three axes of heterogeneity i) \textbf{data composition}, ii) \textbf{resource capacity}, and iii) \textbf{values, culture, and regulation}.  We believe that the fragmented discourse in research areas that study each of these axes individually is a problem, as it reflects their true nature in the real-world.
Furthermore, it is crucial to consider the power dynamics that govern and shape ML. The power in the hands of a selected few entails a one-directional control over the ML systems and the suppression of the unique qualities and perspectives of different communities, in favor of predominantly westernized value systems. Such abuse of power is further exacerbated by the lack of accountability.

Several technical solutions and research efforts aim to promote inclusion and accessibility in ML, however, their common pitfall is the disintegrated view of heterogeneity. Indeed, approaches focusing on democratizing ML often focus on resource heterogeneity and accessibility, while approaches aiming for inclusion consider values as their focal point. 
On a high note, collaborative ML approaches, even in their infancy, are on a promising track toward inclusive and democratic ML. Finally, we call for new ML frameworks that allow for honoring diversity and accessibility.

\begin{acks}
We would like to thank danah boyd for insightful conversation and feedback. This research was partially supported by the Canada CIFAR AI Chair program (Mila), MEI Award (Development for all), and the Natural Sciences and Engineering Research Council of Canada (NSERC). 
\end{acks}


\bibliographystyle{ACM-Reference-Format}
\bibliography{refs.bib}

\end{document}